\pdfoutput=1
\documentclass[10pt,conference]{IEEEtran}

\usepackage[usenames,dvipsnames]{xcolor}
\usepackage{amsmath,amssymb,amsfonts,amsthm}
\usepackage[pdftex]{graphicx}
\usepackage{tabularx}

\usepackage[T1]{fontenc}
\usepackage[utf8]{inputenc}

\usepackage{csquotes}
\usepackage{nicefrac}
\usepackage[textsize=tiny]{todonotes}
\usepackage{booktabs}
\usepackage[hyphens,spaces,obeyspaces]{url}
\usepackage[bookmarks=false]{hyperref}
\usepackage{tikz}
\usetikzlibrary{shapes.geometric, positioning, quantikz}

\makeatletter
\let\MYcaption\@makecaption
\makeatother
\usepackage[font=footnotesize]{subcaption}
\makeatletter
\let\@makecaption\MYcaption
\makeatother

\usepackage[style=ieee, maxnames=6, minnames=1, date=year, doi=false,isbn=false,backend=biber]{biblatex}
\usepackage[binary-units=true, detect-all=true]{siunitx}
\usepackage{etoolbox}
\robustify\bfseries

\usepackage{nicematrix}
\usepackage{flushend}

\newtheorem{example}{Example}

\definecolor{RedEdge}{RGB}{191,40,40}
\definecolor{BlueEdge}{RGB}{40,191,191}
\tikzset{%
	terminal/.style={draw,rectangle,inner sep=2pt,font=\footnotesize,very thick},
	zeronode/.style={fill, draw, circle, minimum width=2pt, inner sep=0pt,color=black},
	qubit/.style={draw,circle,inner sep=0pt,minimum width=0.35cm,minimum height=0.35cm,font=\footnotesize, thin},
	edgeOne/.style={color=RedEdge,ultra thick},
	edgeMOne/.style={color=BlueEdge,ultra thick},
	edgeSqrt/.style={color=RedEdge, thick},
	edgeMSqrt/.style={color=BlueEdge, thick},
}

\newsavebox{\hgate}
\sbox{\hgate}{%
	\begin{tikzpicture}
	\node[qubit] (q0) {$0$};
	\node[terminal, below = 0.2cm of q0] (t) {};
	\draw[edgeSqrt]   ($(q0)+(0,0.4cm)$) -- (q0);
	\draw[edgeOne]  (q0) to[out=-150,in=135] (t);
	\draw[edgeOne]  (q0) to[out=-110,in=110] (t);
	\draw[edgeOne]  (q0) to[out=-70,in=70]   (t);
	\draw[edgeMOne] (q0) to[out=-30,in=45]   (t);
	\end{tikzpicture}
}		

\newsavebox{\czgate}
\sbox{\czgate}{%
	\begin{tikzpicture}
		\matrix[ampersand replacement=\&,column sep={0.3cm,between origins},row sep={0.6cm,between origins}] (qmdd) {
			\& \node[qubit] (q1) {$1$};         \& \\
			\node[qubit] (q0a) {$0$};  \&                             \& \node[qubit] (q0b) {$0$}; \\
			\& \node[terminal] (t) {}; \& \\
		};
		
		\draw[edgeOne] ($(q1)+(0,0.5cm)$) -- (q1);

		\draw[edgeOne] (q1) to[out=-150,in=90]   (q0a);
		\draw[edgeOne] (q1) to[out=-110,in=90] ++(250:0.3cm)  node[zeronode] {};			
		\draw[edgeOne] (q1) to[out=-70,in=90]  ++(290:0.3cm)  node[zeronode] {};			
		\draw[edgeOne] (q1) to[out=-30,in=90]    (q0b);
		
		\draw[edgeOne] (q0a) to[out=-150,in=180] (t);
		\draw[edgeOne] (q0a) to[out=-110,in=90]  ++(250:0.3cm) node[zeronode] {};			
		\draw[edgeOne] (q0a) to[out=-70,in=90]   ++(290:0.3cm)  node[zeronode] {};
		\draw[edgeOne] (q0a) to[out=-30,in=90]   (t);

		\draw[edgeOne] (q0b) to[out=-150,in=90] (t);
		\draw[edgeOne] (q0b) to[out=-110,in=90] ++(250:0.3cm)  node[zeronode] {};
		\draw[edgeOne] (q0b) to[out=-70,in=90]  ++(290:0.3cm)  node[zeronode] {};			
		\draw[edgeMOne] (q0b) to[out=-30,in=0]   (t);
	\end{tikzpicture}
}

\newsavebox{\ddfullstate}
\sbox{\ddfullstate}{%
	\begin{tikzpicture}
		\matrix[ampersand replacement=\&,column sep={0.4cm,between origins},row sep={0.7cm,between origins}] (qmdd) {
			\&\&\& \node[qubit] (q3) {$3$}; \&\&\& \\
			\& \node[qubit] (q2a) {$2$}; \&\&\&\& \node[qubit] (q2b) {$2$}; \& \\
			\node[qubit] (q1a) {$1$}; \&\& \node[qubit] (q1b) {$1$}; \&\& \node[qubit] (q1c) {$1$}; \&\& \node[qubit] (q1d) {$1$}; \\
			\& \node[qubit] (q0a) {$0$}; \&\&\&\& \node[qubit] (q0b) {$0$}; \& \\
			\&\&\&  \node[terminal] (t) {}; \&\&\& \\
		};
		
		\draw[edgeOne] ($(q3)+(0,0.55cm)$) -- (q3);
		
		\draw[edgeSqrt] (q3) to[out=-150,in=90] (q2a);			
		\draw[edgeSqrt] (q3) to[out=-30,in=90] (q2b);
		
		\draw[edgeSqrt] (q2a) to[out=-150,in=90] (q1a);			
		\draw[edgeSqrt] (q2a) to[out=-30,in=120] (q1c);
		
		\draw[edgeSqrt] (q2b) to[out=-150,in=60] (q1b);			
		\draw[edgeSqrt] (q2b) to[out=-30,in=90] (q1d);
		
		\draw[edgeSqrt] (q1a) to[out=-150,in=140] (q0a);			
		\draw[edgeSqrt] (q1a) to[out=-30,in=100] (q0a);
		
		\draw[edgeSqrt] (q1b) to[out=-150,in=80] (q0a);			
		\draw[edgeMSqrt] (q1b) to[out=-30,in=40] (q0a);
		
		\draw[edgeSqrt] (q1c) to[out=-150,in=140] (q0b);			
		\draw[edgeSqrt] (q1c) to[out=-30,in=100]  (q0b);
		
		\draw[edgeSqrt] (q1d) to[out=-150,in=80] (q0b);			
		\draw[edgeMSqrt] (q1d) to[out=-30,in=40]  (q0b);
		
		\draw[edgeSqrt] (q0a) to[out=-150,in=160] (t);			
		\draw[edgeSqrt] (q0a) to[out=-30,in=100]  (t);
		
		\draw[edgeSqrt] (q0b) to[out=-150,in=80] (t);			
		\draw[edgeMSqrt] (q0b) to[out=-30,in=20]  (t);
	\end{tikzpicture}
}

\newsavebox{\legend}
\sbox{\legend}{%
	\begin{tikzpicture}
	\draw[edgeOne]   (0, 0) node[above,xshift=0.5cm] {$1$}  -- (1, 0);
	\draw[edgeMOne]  (1.5, 0) node[above,xshift=0.45cm] {$-1$}  -- (2.5, 0);
	\draw[edgeSqrt]   (3, 0) node[above,xshift=0.5cm] {$\tfrac{1}{\sqrt{2}}$}  -- (4, 0);
	\draw[edgeMSqrt]  (4.5, 0) node[above,xshift=0.45cm] {$-\tfrac{1}{\sqrt{2}}$}  -- (5.5, 0);
	\end{tikzpicture}
}

\newsavebox{\pzero}
\sbox{\pzero}{%
	\begin{tikzpicture}
	\node[qubit] (q0) {$0$};
	\node[below = 0.3cm of q0, terminal] (t) {};
	\draw[edgeOne] ($(q0)+(0,0.45cm)$) -- (q0);
	\draw[edgeOne] (q0) to[out=-150,in=135] (t);
	\draw[edgeOne] (q0) to[out=-110,in=90] ++(250:0.3cm)node[zeronode] {};			
	\draw[edgeOne] (q0) to[out=-70,in=90] ++(290:0.3cm)node[zeronode] {};		
	\draw[edgeOne] (q0) to[out=-20,in=90] ++(330:0.35cm) node[zeronode] {};
	\end{tikzpicture}	
}

\newsavebox{\pone}
\sbox{\pone}{%
	\begin{tikzpicture}
	\node[qubit] (q0) {$0$};
	\node[below = 0.3cm of q0, terminal] (t) {};
	\draw[edgeOne] ($(q0)+(0,0.45cm)$) -- (q0);
	\draw[edgeOne] (q0) to[out=-150,in=135] ++(210:0.35cm) node[zeronode] {};
	\draw[edgeOne] (q0) to[out=-110,in=90] ++(250:0.3cm)node[zeronode] {};			
	\draw[edgeOne] (q0) to[out=-70,in=90] ++(290:0.3cm)node[zeronode] {};		
	\draw[edgeOne] (q0) to[out=-20,in=45] (t);
	\end{tikzpicture}	
}

\newsavebox{\identity}
\sbox{\identity}{%
	\begin{tikzpicture}
	\node[qubit] (q0) {$0$};
	\node[below = 0.3cm of q0, terminal] (t) {};
	\draw[edgeOne] ($(q0)+(0,0.45cm)$) -- (q0);
	\draw[edgeOne] (q0) to[out=-150,in=135] (t);
	\draw[edgeOne] (q0) to[out=-110,in=90] ++(250:0.3cm)node[zeronode] {};			
	\draw[edgeOne] (q0) to[out=-70,in=90] ++(290:0.3cm)node[zeronode] {};		
	\draw[edgeOne] (q0) to[out=-20,in=45] (t);
	\end{tikzpicture}	
}

\newsavebox{\z}
\sbox{\z}{%
	\begin{tikzpicture}
	\node[qubit] (q0) {$0$};
	\node[below = 0.3cm of q0, terminal] (t) {};
	\draw[edgeOne] ($(q0)+(0,0.45cm)$) -- (q0);
	\draw[edgeOne] (q0) to[out=-150,in=135] (t);
	\draw[edgeOne] (q0) to[out=-110,in=90] ++(250:0.3cm)node[zeronode] {};			
	\draw[edgeOne] (q0) to[out=-70,in=90] ++(290:0.3cm)node[zeronode] {};		
	\draw[edgeMOne] (q0) to[out=-20,in=45] (t);
	\end{tikzpicture}	
}

\newsavebox{\dda}
\sbox{\dda}{%
	\begin{tikzpicture}
	\node[qubit] (q3) {$3$};
	\node[below = 0.35cm of q3, qubit] (q2) {$2$};
	\node[below = 0.35cm of q2, terminal] (t) {};
	
	\draw[edgeOne] ($(q3)+(0,0.55cm)$) -- (q3);
	\draw[edgeSqrt] (q3) to[out=-150,in=120] (q2);		
	\draw[edgeSqrt] (q3) to[out=-20,in=90] ++(330:0.3cm) node[zeronode] {};
	\draw[edgeSqrt] (q2) to[out=-150,in=120] (t);	
	\draw[edgeSqrt] (q2) to[out=-20,in=90] ++(330:0.3cm) node[zeronode] {};
	\end{tikzpicture}	
}

\newsavebox{\ddb}
\sbox{\ddb}{%
	\begin{tikzpicture}
	\node[qubit] (q3) {$1$};
	\node[below = 0.35cm of q3, qubit] (q2) {$0$};
	\node[below = 0.35cm of q2, terminal] (t) {};
	
	\draw[edgeOne] ($(q3)+(0,0.55cm)$) -- (q3);
	\draw[edgeSqrt] (q3) to[out=-150,in=120] (q2);		
	\draw[edgeSqrt] (q3) to[out=-20,in=60] (q2);
	\draw[edgeSqrt] (q2) to[out=-150,in=120] (t);	
	\draw[edgeSqrt] (q2) to[out=-20,in=60] (t);
	\end{tikzpicture}	
}

\newsavebox{\ddc}
\sbox{\ddc}{%
	\begin{tikzpicture}
	\node[qubit] (q3) {$3$};
	\node[below = 0.35cm of q3, qubit] (q2) {$2$};
	\node[below = 0.35cm of q2, terminal] (t) {};
	
	\draw[edgeOne] ($(q3)+(0,0.55cm)$) -- (q3);
	\draw[edgeSqrt] (q3) to[out=-150,in=120] (q2);		
	\draw[edgeSqrt] (q3) to[out=-20,in=90] ++(330:0.3cm) node[zeronode] {};
	\draw[edgeSqrt] (q2) to[out=-150,in=90] ++(210:0.3cm) node[zeronode] {};
	\draw[edgeSqrt] (q2) to[out=-20,in=60] (t);
	\end{tikzpicture}	
}

\newsavebox{\ddd}
\sbox{\ddd}{%
	\begin{tikzpicture}
	\node[qubit] (q3) {$1$};
	\node[below = 0.35cm of q3, qubit] (q2) {$0$};
	\node[below = 0.35cm of q2, terminal] (t) {};
	
	\draw[edgeOne] ($(q3)+(0,0.55cm)$) -- (q3);
	\draw[edgeSqrt] (q3) to[out=-150,in=120] (q2);		
	\draw[edgeSqrt] (q3) to[out=-20,in=60] (q2);
	\draw[edgeSqrt] (q2) to[out=-150,in=120] (t);	
	\draw[edgeMSqrt] (q2) to[out=-20,in=60] (t);
	\end{tikzpicture}	
}

\newsavebox{\dde}
\sbox{\dde}{%
	\begin{tikzpicture}
	\node[qubit] (q3) {$3$};
	\node[below = 0.35cm of q3, qubit] (q2) {$2$};
	\node[below = 0.35cm of q2, terminal] (t) {};
	
	\draw[edgeOne] ($(q3)+(0,0.55cm)$) -- (q3);
	\draw[edgeSqrt] (q3) to[out=-150,in=90]  ++(210:0.3cm) node[zeronode] {};		
	\draw[edgeSqrt] (q3) to[out=-20,in=60] (q2);
	\draw[edgeSqrt] (q2) to[out=-150,in=120] (t);	
	\draw[edgeSqrt] (q2) to[out=-20,in=90] ++(330:0.3cm) node[zeronode] {};
	\end{tikzpicture}	
}

\newsavebox{\ddf}
\sbox{\ddf}{%
	\begin{tikzpicture}
	\node[qubit] (q3) {$1$};
	\node[below = 0.35cm of q3, qubit] (q2) {$0$};
	\node[below = 0.35cm of q2, terminal] (t) {};
	
	\draw[edgeOne] ($(q3)+(0,0.55cm)$) -- (q3);
	\draw[edgeSqrt] (q3) to[out=-150,in=120] (q2);		
	\draw[edgeMSqrt] (q3) to[out=-20,in=60] (q2);
	\draw[edgeSqrt] (q2) to[out=-150,in=120] (t);	
	\draw[edgeSqrt] (q2) to[out=-20,in=60] (t);
	\end{tikzpicture}	
}

\newsavebox{\ddg}
\sbox{\ddg}{%
	\begin{tikzpicture}
	\node[qubit] (q3) {$3$};
	\node[below = 0.35cm of q3, qubit] (q2) {$2$};
	\node[below = 0.35cm of q2, terminal] (t) {};
	
	\draw[edgeOne] ($(q3)+(0,0.55cm)$) -- (q3);
	\draw[edgeSqrt] (q3) to[out=-150,in=90]  ++(210:0.3cm) node[zeronode] {};		
	\draw[edgeSqrt] (q3) to[out=-20,in=60] (q2);
	\draw[edgeSqrt] (q2) to[out=-150,in=90] ++(210:0.3cm) node[zeronode] {};
	\draw[edgeSqrt] (q2) to[out=-20,in=60] (t);
	\end{tikzpicture}	
}

\newsavebox{\ddh}
\sbox{\ddh}{%
	\begin{tikzpicture}
	\node[qubit] (q3) {$1$};
	\node[below = 0.35cm of q3, qubit] (q2) {$0$};
	\node[below = 0.35cm of q2, terminal] (t) {};
	
	\draw[edgeOne] ($(q3)+(0,0.55cm)$) -- (q3);
	\draw[edgeSqrt] (q3) to[out=-150,in=120] (q2);		
	\draw[edgeMSqrt] (q3) to[out=-20,in=60] (q2);
	\draw[edgeSqrt] (q2) to[out=-150,in=120] (t);	
	\draw[edgeMSqrt] (q2) to[out=-20,in=60] (t);
	\end{tikzpicture}	
}

\newsavebox{\ddtensora}
\sbox{\ddtensora}{%
	\begin{tikzpicture}
	\node[qubit] (q3) {$3$};
	\node[below = 0.35cm of q3, qubit] (q2) {$2$};
	\node[below = 0.35cm of q2, qubit] (q1) {$1$};
	\node[below = 0.35cm of q1, qubit] (q0) {$0$};
	\node[below = 0.35cm of q0, terminal] (t) {};
	
	\draw[edgeOne] ($(q3)+(0,0.55cm)$) -- (q3);
	\draw[edgeSqrt] (q3) to[out=-150,in=120] (q2);		
	\draw[edgeSqrt] (q3) to[out=-20,in=90] ++(330:0.3cm) node[zeronode] {};
	\draw[edgeSqrt] (q2) to[out=-150,in=120] (q1);	
	\draw[edgeSqrt] (q2) to[out=-20,in=90] ++(330:0.3cm) node[zeronode] {};
	\draw[edgeSqrt] (q1) to[out=-150,in=120] (q0);		
	\draw[edgeSqrt] (q1) to[out=-20,in=60] (q0);
	\draw[edgeSqrt] (q0) to[out=-150,in=120] (t);	
	\draw[edgeSqrt] (q0) to[out=-20,in=60] (t);
	\end{tikzpicture}	
}

\newsavebox{\ddtensorb}
\sbox{\ddtensorb}{%
	\begin{tikzpicture}
	\node[qubit] (q3) {$3$};
	\node[below = 0.35cm of q3, qubit] (q2) {$2$};
	\node[below = 0.35cm of q2, qubit] (q1) {$1$};
	\node[below = 0.35cm of q1, qubit] (q0) {$0$};
	\node[below = 0.35cm of q0, terminal] (t) {};
	
	\draw[edgeOne] ($(q3)+(0,0.55cm)$) -- (q3);
	\draw[edgeSqrt] (q3) to[out=-150,in=120] (q2);		
	\draw[edgeSqrt] (q3) to[out=-20,in=90] ++(330:0.3cm) node[zeronode] {};
	\draw[edgeSqrt] (q2) to[out=-150,in=90] ++(210:0.3cm) node[zeronode] {};
	\draw[edgeSqrt] (q2) to[out=-20,in=60] (q1);
	\draw[edgeSqrt] (q1) to[out=-150,in=120] (q0);		
	\draw[edgeSqrt] (q1) to[out=-20,in=60] (q0);
	\draw[edgeSqrt] (q0) to[out=-150,in=120] (t);	
	\draw[edgeMSqrt] (q0) to[out=-20,in=60] (t);
	\end{tikzpicture}	
}

\newsavebox{\ddtensorc}
\sbox{\ddtensorc}{%
	\begin{tikzpicture}
	\node[qubit] (q3) {$3$};
	\node[below = 0.35cm of q3, qubit] (q2) {$2$};
	\node[below = 0.35cm of q2, qubit] (q1) {$1$};
	\node[below = 0.35cm of q1, qubit] (q0) {$0$};
	\node[below = 0.35cm of q0, terminal] (t) {};
	
	\draw[edgeOne] ($(q3)+(0,0.55cm)$) -- (q3);
	\draw[edgeSqrt] (q3) to[out=-150,in=90]  ++(210:0.3cm) node[zeronode] {};		
	\draw[edgeSqrt] (q3) to[out=-20,in=60] (q2);
	\draw[edgeSqrt] (q2) to[out=-150,in=120] (q1);	
	\draw[edgeSqrt] (q2) to[out=-20,in=90] ++(330:0.3cm) node[zeronode] {};
	\draw[edgeSqrt] (q1) to[out=-150,in=120] (q0);		
	\draw[edgeMSqrt] (q1) to[out=-20,in=60] (q0);
	\draw[edgeSqrt] (q0) to[out=-150,in=120] (t);	
	\draw[edgeSqrt] (q0) to[out=-20,in=60] (t);
	\end{tikzpicture}	
}

\newsavebox{\ddtensord}
\sbox{\ddtensord}{%
	\begin{tikzpicture}
	\node[qubit] (q3) {$3$};
	\node[below = 0.35cm of q3, qubit] (q2) {$2$};
	\node[below = 0.35cm of q2, qubit] (q1) {$1$};
	\node[below = 0.35cm of q1, qubit] (q0) {$0$};
	\node[below = 0.35cm of q0, terminal] (t) {};
	
	\draw[edgeOne] ($(q3)+(0,0.55cm)$) -- (q3);
	\draw[edgeSqrt] (q3) to[out=-150,in=90]  ++(210:0.3cm) node[zeronode] {};		
	\draw[edgeSqrt] (q3) to[out=-20,in=60] (q2);
	\draw[edgeSqrt] (q2) to[out=-150,in=90] ++(210:0.3cm) node[zeronode] {};
	\draw[edgeSqrt] (q2) to[out=-20,in=60] (q1);
	\draw[edgeSqrt] (q1) to[out=-150,in=120] (q0);		
	\draw[edgeMSqrt] (q1) to[out=-20,in=60] (q0);
	\draw[edgeSqrt] (q0) to[out=-150,in=120] (t);	
	\draw[edgeMSqrt] (q0) to[out=-20,in=60] (t);
	\end{tikzpicture}	
}

\newsavebox{\ddtensoradda}
\sbox{\ddtensoradda}{%
	\begin{tikzpicture}
	\matrix[ampersand replacement=\&,column sep={0.4cm,between origins},row sep={0.7cm,between origins}] (qmdd) {
		                          \& \node[qubit] (q3) {$3$}; \& \\
	    \node[qubit] (q2a) {$2$}; \&                          \& \node[qubit] (q2b) {$2$}; \\
		\node[qubit] (q1a) {$1$}; \&                           \& \node[qubit] (q1b) {$1$};  \\
                                  \& \node[qubit] (q0) {$0$}; \&\\
		\& \node[terminal] (t) {}; \& \\
	};
	
	\draw[edgeOne] ($(q3)+(0,0.55cm)$) -- (q3);
	
	\draw[edgeSqrt] (q3) to[out=-150,in=90] (q2a);			
	\draw[edgeSqrt] (q3) to[out=-30,in=90] (q2b);
	
	\draw[edgeSqrt] (q2a) to[out=-150,in=90] (q1a);			
	\draw[edgeSqrt] (q2a) to[out=-30,in=90] ++(330:0.3cm) node[zeronode] {};
	
	\draw[edgeSqrt] (q2b) to[out=-150,in=90] (q1b);			
	\draw[edgeSqrt] (q2b) to[out=-30,in=90] ++(330:0.3cm) node[zeronode] {};
	
	\draw[edgeSqrt] (q1a) to[out=-150,in=140] (q0);			
	\draw[edgeSqrt] (q1a) to[out=-30,in=100] (q0);
	
	\draw[edgeSqrt] (q1b) to[out=-150,in=80] (q0);			
	\draw[edgeMSqrt] (q1b) to[out=-30,in=40] (q0);
	
	\draw[edgeSqrt] (q0) to[out=-150,in=100] (t);			
	\draw[edgeSqrt] (q0) to[out=-30,in=80]  (t);
	\end{tikzpicture}
}

\newsavebox{\ddtensoraddb}
\sbox{\ddtensoraddb}{%
	\begin{tikzpicture}
	\matrix[ampersand replacement=\&,column sep={0.4cm,between origins},row sep={0.7cm,between origins}] (qmdd) {
		\& \node[qubit] (q3) {$3$}; \& \\
		\node[qubit] (q2a) {$2$}; \&                          \& \node[qubit] (q2b) {$2$}; \\
		\node[qubit] (q1a) {$1$}; \&                           \& \node[qubit] (q1b) {$1$};  \\
		\& \node[qubit] (q0) {$0$}; \&\\
		\& \node[terminal] (t) {}; \& \\
	};
	
	\draw[edgeOne] ($(q3)+(0,0.55cm)$) -- (q3);
	
	\draw[edgeSqrt] (q3) to[out=-150,in=90] (q2a);			
	\draw[edgeSqrt] (q3) to[out=-30,in=90] (q2b);
	
	\draw[edgeSqrt] (q2a) to[out=-150,in=90] ++(210:0.3cm) node[zeronode] {};
	\draw[edgeSqrt] (q2a) to[out=-30,in=90](q1a);			
	
	\draw[edgeSqrt] (q2b) to[out=-150,in=90] ++(210:0.3cm) node[zeronode] {};	
	\draw[edgeSqrt] (q2b) to[out=-30,in=90](q1b);	
	
	\draw[edgeSqrt] (q1a) to[out=-150,in=140] (q0);			
	\draw[edgeSqrt] (q1a) to[out=-30,in=100] (q0);
	
	\draw[edgeSqrt] (q1b) to[out=-150,in=80] (q0);			
	\draw[edgeMSqrt] (q1b) to[out=-30,in=40] (q0);
	
	\draw[edgeSqrt] (q0) to[out=-150,in=100] (t);			
	\draw[edgeMSqrt] (q0) to[out=-30,in=80]  (t);
	\end{tikzpicture}
}

\begin{document}

\title{Hybrid Schrödinger-Feynman Simulation\\of Quantum Circuits With Decision Diagrams}

\author{
	\IEEEauthorblockN{Lukas Burgholzer\IEEEauthorrefmark{1}\hspace*{1.5cm}Hartwig Bauer\IEEEauthorrefmark{1}\hspace*{1.5cm}Robert Wille\IEEEauthorrefmark{1}\IEEEauthorrefmark{2}}
	\IEEEauthorblockA{\IEEEauthorrefmark{1}Institute for Integrated Circuits, Johannes Kepler University Linz, Austria}
	\IEEEauthorblockA{\IEEEauthorrefmark{2}Software Competence Center Hagenberg GmbH (SCCH), Austria}
	\IEEEauthorblockA{\href{mailto:lukas.burgholzer@jku.at}{lukas.burgholzer@jku.at}\hspace{1.5cm}\href{mailto:hartwig.bauer@jku.at}{hartwig.bauer@jku.at}\hspace{1.5cm} \href{mailto:robert.wille@jku.at}{robert.wille@jku.at}\\
	\url{https://iic.jku.at/eda/research/quantum/}}
	\vspace*{-2.1em}
}

\maketitle

\begin{abstract}
Classical simulations of quantum computations are vital for the future development of this emerging technology.
To this end, decision diagrams have been proposed as a complementary technique which frequently allows to tackle the inherent exponential complexity of these simulations.
In the worst case, however, they still cannot escape this complexity. Additionally, while other techniques make use of all the available processing power, decision diagram-based simulation to date cannot exploit the many processing units of today's systems.
In this work, we show that both problems can be tackled together by employing a hybrid Schrödinger-Feynman scheme for the simulation.
More precisely, we show that realizing such a scheme with decision diagrams is indeed possible, we discuss the resulting problems in its 
realization, and propose solutions how they can be handled.
Experimental evaluations confirm that this significantly advances the state of the art in decision diagram-based simulation---allowing to simulate certain hard circuits within minutes that could not be simulated in a whole day thus far. 
\end{abstract}


\section{Introduction}
\label{sec:introduction}

Despite actual quantum computers being available on the cloud, the simulation of quantum circuits on classical machines remains crucial for the development and design of future quantum computing applications. 
Such simulations provide insights into the inner workings of a quantum system and allow, e.g., to analyze quantum algorithms or verify the output of physical quantum computers.
To this end, several notions of what such a classical simulation entails exist.
In this work, we consider \emph{strong} simulation of quantum circuits, i.e., we want to compute \emph{all} complex amplitudes of the quantum state resulting from the execution of the circuit (as, e.g., in~\cite{hanerPetabyteSimulation45Qubit2017,doiQuantumComputingSimulator2019,jonesQuESTHighPerformance2018,guerreschiIntelQuantumSimulator2020, wuFullstateQuantumCircuit2019, zulehnerAdvancedSimulationQuantum2019, samoladasImprovedBDDAlgorithms2008, viamontesHighperformanceQuIDDBasedSimulation2004, zulehnerMatrixVectorVsMatrixMatrix2019, grurlArraysVsDecision2020, hillmichAccurateNeededEfficient2020, aaronsonComplexityTheoreticFoundationsQuantum2016, markovQuantumSupremacyBoth2018, chen64QubitQuantumCircuit2018, pednaultParetoefficientQuantumCircuit2020}).
This is in contrast to methods based, e.g., on tensor networks that typically only compute individual amplitudes or a small batch of them~\cite{markovSimulatingQuantumComputation2008, boixoSimulationLowdepthQuantum2018, boixoCharacterizingQuantumSupremacy2018, villalongaFlexibleHighperformanceSimulator2019, huangClassicalSimulationQuantum2020, grayHyperoptimizedTensorNetwork2021, panSimulatingSycamoreQuantum2021}.
Furthermore, we do not consider \emph{approximate} simulation, which trades computational complexity for accuracy of the result (as, e.g., in~\cite{hillmichAccurateNeededEfficient2020, markovQuantumSupremacyBoth2018}).

The most fundamental technique of simulating a quantum circuit is called \emph{Schrödinger-style} simulation.
There, a complete representation of the quantum system's state is stored and manipulated throughout the computation.
While \mbox{straight-forward} in principle, this quickly amounts to a complex task, due to the underlying representation requiring the storage and manipulation of $2^n$ complex amplitudes for an $n$-qubit system.

While this complexity is frequently tackled by massively parallel computations on \emph{arrays}~\mbox{\cite{hanerPetabyteSimulation45Qubit2017,doiQuantumComputingSimulator2019,jonesQuESTHighPerformance2018,guerreschiIntelQuantumSimulator2020, wuFullstateQuantumCircuit2019}} using supercomputer clusters with immense amounts of memory and processing power,
\emph{decision diagrams}~\cite{chin-yungExtendedXQDDRepresentation2011,millerQMDDDecisionDiagram2006, zulehnerHowEfficientlyHandle2019,hongTensorNetworkBased2020} have been proposed as a complementary technique that aims at compactly representing and efficiently manipulating the $2^n$ complex amplitudes and, hence, often allows to conduct corresponding simulations on just a single desktop computer~\mbox{\cite{zulehnerAdvancedSimulationQuantum2019, samoladasImprovedBDDAlgorithms2008, viamontesHighperformanceQuIDDBasedSimulation2004, hillmichAccurateNeededEfficient2020, zulehnerMatrixVectorVsMatrixMatrix2019,grurlArraysVsDecision2020}}.
But, in the worst case, also decision diagrams are subject to the inherent exponential complexity. 

Complementary to that, there exists another formalism suited for simulating quantum circuits which reduces the memory complexity of the simulation by breaking it down into simpler parts.
This is known as \emph{Feynman-style} path summation~\cite{bernsteinQuantumComplexityTheory1997,aaronsonComplexityTheoreticFoundationsQuantum2016}.
From a high-level point of view, each gate connecting two or more qubits in a quantum circuit introduces a decision point from which the simulation branches (this notion will be made more precise later in \autoref{sec:slicing}). 
As the name implies, \emph{Feynman-style} path summation calculates the result of each path and sums up all the individual contributions---resulting in the final quantum state. Since the number of paths is exponential depending on the number of decision points, this approach requires exponential runtime (but usually avoids too harsh memory requirements).

Over the recent years, mixtures of both schemes emerged~\cite{aaronsonComplexityTheoreticFoundationsQuantum2016, chen64QubitQuantumCircuit2018, markovQuantumSupremacyBoth2018, pednaultParetoefficientQuantumCircuit2020}---often referred to as \emph{hybrid \mbox{Schrödinger-Feynman} simulation}.
These approaches strive to use all the available memory and processing units in order to efficiently simulate quantum circuits which would (1)~run into memory bottlenecks using Schrödinger-style simulation, or (2)~take exceedingly long using \mbox{Feynman-style} path summation---eventually trading-off the respective memory and runtime requirements. 
However, while this hybrid scheme can easily be realized, e.g., using arrays, no solution for decision diagrams exists yet (in fact, there are even discussions that a corresponding realization might not be possible at all~\cite{hongTensorNetworkBased2020}). This constitutes a severe drawback as it keeps decision diagram-based simulation stuck with Schrödinger-style simulation that is only suitable if the compact representations of decision diagrams allow to escape the exponential memory requirements (while, in all remaining cases, decision diagrams even impose a severe overhead compared to rather simple arrays).

\pagebreak

In this work, we show that realizing a \emph{hybrid Schrödinger-Feynman} scheme with decision diagrams is indeed possible---even if some problems arise when doing so.
We describe a possible realization, 
discuss what problems exactly arise, and propose solutions to overcome them.
Eventually, the first hybrid Schrödinger-Feynman quantum circuit simulation approach results which works with decision diagrams. 
Experimental evaluations confirm that this significantly advances the state of the art in decision diagram-based simulation---allowing to simulate certain hard circuits within minutes that could not be simulated in a whole day thus far.
An implementation of the proposed simulation technique is publicly available at \url{https://github.com/iic-jku/ddsim}.

The remainder of this paper is structured as follows. \autoref{sec:background} reviews the basics of (decision diagram-based) quantum circuit simulation.
Then, \autoref{sec:general_idea} describes the general idea of the hybrid Schrödinger-Feynman technique. 
\autoref{sec:slicing_dds} describes the realization of such techniques for decision diagrams, the resulting limitations, and how they can be handled. 
Afterwards, \autoref{sec:results} shows the experimental results before the paper is concluded in \autoref{sec:conclusions}.

\section{Background}
\label{sec:background}
To keep this paper self-contained, this section briefly reviews the basics of quantum circuits and their simulation as well as how this simulation can be conducted using decision diagrams.
We refer the interested reader to~\cite{nielsenQuantumComputationQuantum2010, zulehnerAdvancedSimulationQuantum2019} for a more detailed overview of either topic. Furthermore, an online visualization tool, which makes decision diagrams for quantum computing more accessible~\cite{willeVisualizingDecisionDiagrams2021}, is available at \url{https://iic.jku.at/eda/research/quantum_dd/tool/}.

\subsection{Quantum Circuits and Their Simulation}
In general, the state $\ket{\varphi}$ of an $n$-qubit quantum system is described as a linear combination of $2^n$ basis states, i.e., 
\[
\ket{\varphi} = \sum_{x\in\{0,1\}^n} \alpha_x \ket{x} \mbox{ with } \alpha_x\in\mathbb{C} \mbox{ and }\sum_{x\in\{0,1\}^n} \vert\alpha_x\vert^2 = 1.
\]
This is frequently represented as $[\alpha_{0\dots 0}, \dots, \alpha_{1\dots 1}]^\top$ and referred to as \emph{state vector}.
Measuring the state of a quantum system probabilistically collapses the system's state to one of the basis states---each with probability $\vert \alpha_x\vert^2$ for $x\in\{0,1\}^n$.

\begin{example}\label{ex:state}
Consider a quantum system consisting of four qubits that resides in the all-zero initial state $\ket{0000}$. 
The system's state vector consists of $2^4=16$ complex amplitudes and is represented by 
\[
[1,0,0,0,0,0,0,0,0,0,0,0,0,0,0,0]^\top.
\]
\end{example}

The state of a quantum system is manipulated by quantum operations (frequently called \emph{quantum gates}).
Any $k$-qubit quantum gate can be identified with a unitary matrix~$U$ of size $2^k\times 2^k$, which can be extended to the whole system size by forming tensor products with identity matrices.
A quantum circuit describes a sequence of gates, that are applied to a quantum system.
Applying a quantum gate $U$ to the state~$\ket{\varphi}$ of a quantum system corresponds to extending its matrix representation to the system size and computing the matrix-vector product of the resulting matrix with the state vector---yielding a new state vector $\ket{\varphi'}$.
Thus, simulating a quantum circuit entails the successive application of all individual gates to the initial state of a quantum system in order to obtain the final state (vector). An example illustrates the idea:

\begin{figure}[t]
	\centering
	\resizebox{0.99\linewidth}{!}{
	\begin{tikzpicture}
	\node[scale=0.8] (G) {
	\begin{quantikz}[column sep=10pt, row sep={17.5pt,between origins}]
		\lstick{$q_3\colon\ket{0}$} & \gate[style={fill=RoyalBlue!20}]{H}  & \ctrl{2} & \qw      & \qw \\
		\lstick{$q_2\colon\ket{0}$} & \gate[style={fill=RoyalBlue!20}]{H}  & \qw & \ctrl{2} & \qw \\
		\lstick{$q_1\colon\ket{0}$} & \gate[style={fill=RoyalBlue!20}]{H}  & \gate[style={fill=green!20}]{Z} & \qw      & \qw \\
		\lstick{$q_0\colon\ket{0}$} & \gate[style={fill=RoyalBlue!20}]{H}  & \qw      & \gate[style={fill=green!20}]{Z} & \qw
	\end{quantikz}
	};
	
	\node[left = 2cm of G.165] (H) {
	\begin{quantikz}[column sep=10pt, row sep={17.5pt,between origins}, scale=0.95]
		& \gate[style={fill=RoyalBlue!20}]{H}  & \qw
	\end{quantikz}
	};	

	\node[right = 1.6cm of H.west] (Hmat) {
	$=\frac{1}{\sqrt{2}}\begin{bNiceMatrix}[small]1&1\\1&-1\end{bNiceMatrix}$
	};
	
	\node[below = 1cm of H.north] (cZ) {
	\begin{quantikz}[column sep=10pt, row sep={17.5pt,between origins}, scale=0.95]
		& \ctrl{1}  & \qw \\
		& \gate[style={fill=green!20}]{Z} & \qw
	\end{quantikz}
	};	

	\node[right = 1.6cm of cZ.west] (cZmat) {
	$=\begin{bNiceMatrix}[small]1 & & & \\ & 1 & & \\ & & 1 & \\ & & & -1\end{bNiceMatrix}$
	};
	\end{tikzpicture}}
	\caption{Quantum gates and quantum circuit $G$}
	\label{fig:circuit}	
\end{figure}
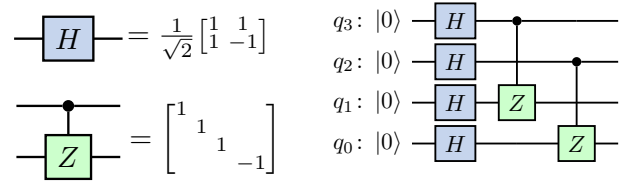

\begin{example}\label{ex:circuit}
An example of a single-qubit gate (the Hadamard) and a two-qubit gate (the controlled-Z operation) is shown on the left-hand side of \autoref{fig:circuit} with their corresponding matrix representations.
The right-hand side of \autoref{fig:circuit} shows a quantum circuit~$G$ acting on four qubits that uses these gates. 
Simulating this circuit with the input state $\ket{0000}$ (see \autoref{ex:state}) results in the final state vector:
	\[
	\tfrac{1}{4}* [1,1,1,1,1,-1,1,-1,1,1,-1,-1,1,-1,-1,1]^\top.
	\]
\end{example}

\subsection{Decision Diagram-based Simulation}\label{sec:ddsim}

The size of the state vector is inherently exponential with respect to the number of qubits.
Not only does this always incur an exponential memory footprint, but it also implies that all operations have to be conducted on an exponentially large data structure.

\emph{Decision diagrams}~\cite{zulehnerHowEfficientlyHandle2019,hongTensorNetworkBased2020,chin-yungExtendedXQDDRepresentation2011,millerQMDDDecisionDiagram2006} have been proposed 
as a complementary approach for efficiently representing and manipulating the state of a quantum system by exploiting redundancies in the underlying representation.
They represent quantum states and operations as weighted, directed, acyclic graphs.
To this end, a given state vector with its complex amplitudes \(\alpha_i\) for \(i \in \{0,1\}^n\) is decomposed into sub-vectors
\begin{gather*}
        [\alpha_{00\ldots 0}, \hdots, \alpha_{1\ldots 1}]^\top \\
        [\alpha_{0x}]^\top \qquad\qquad\qquad [\alpha_{1x}]^\top \\
        [\alpha_{00y}]^\top \quad [\alpha_{01y}]^\top \qquad [\alpha_{10y}]^\top \quad [\alpha_{11y}]^\top
\end{gather*}
with $x\in\{0,1\}^{n-1}$ and $y\in\{0,1\}^{n-2}$, until only individual amplitudes remain.
The resulting decision diagram has $n$ levels of nodes (labeled ${n-1}$ to $0$) where each node \(i\) has exactly two successors---indicating whether the corresponding path leads to an amplitude where qubit $q_i$ is in the state \ket{0} or \ket{1}.
During these decompositions, common factors are extracted as edge weights according to a normalization scheme and equivalent \mbox{sub-vectors} are represented by the same node---allowing to exploit potential redundancies in the representation.
The amplitude of a given basis state can then be reconstructed from the multiplication of the edge weights along the path from the root node to the terminal node.

\pagebreak

\begin{figure}[t]
	\centering
		\begin{tikzpicture}
		
		\node[yshift=0.75cm] (H) {
			\begin{quantikz}[column sep=10pt, row sep={17.5pt,between origins}, scale=0.95]
			& \gate[style={fill=RoyalBlue!20}]{H}  & \qw
			\end{quantikz}
		};	
	
		\node[right = 1.6cm of H.west] (EqH) {
			$=$
		};
		
		\node[right = 1.9cm of H.west] (Hmat) {
			\usebox{\hgate}
		};
		
		\node[below = 2cm of H.north, xshift=-0.5] (cZ) {
			\begin{quantikz}[column sep=10pt, row sep={17.5pt,between origins}, scale=0.95]
			& \ctrl{1}  & \qw \\
			& \gate[style={fill=green!20}]{Z} & \qw
			\end{quantikz}
		};	
		
		\node[right = 1.6cm of cZ.west] (EqCZ) {
			$=$
		};
	
		\node[right = 0.2cm of cZ.15, scale=0.88] (cZmat) {
			\usebox{\czgate}
		};

		\node[left = 2cm of H.0, yshift=-1cm] (G) {
			\usebox{\ddfullstate}
		};
		\node [below= 0.1mm of G.south,inner sep=0pt]{\parbox{0.4\linewidth}{\subcaption{Representing final state from \autoref{ex:circuit}}\label{fig:ddstate}}};
		\node [below= 5.2mm of EqCZ.south,inner sep=0pt]{\parbox{0.4\linewidth}{\subcaption{Representing operations from \autoref{fig:circuit}}\label{fig:ddop}}};

		\node[above = 0.01 cm of G, xshift=2cm] (Legend) {
			\usebox{\legend}
		};
		\end{tikzpicture}
	\caption{Decision diagrams}
	\label{fig:dds}
\end{figure}
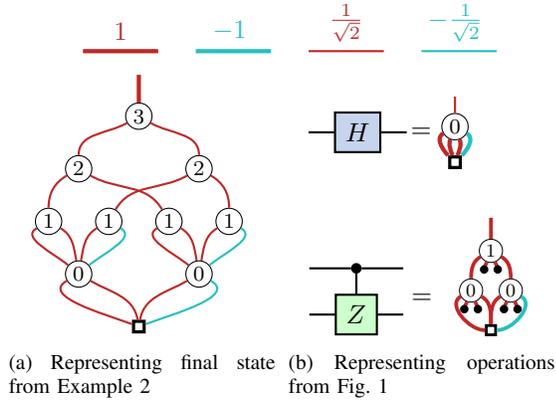

\begin{example}\label{ex:statedd}
	Consider again the final quantum state from \autoref{ex:circuit}.
	The corresponding decision diagram is shown in \autoref{fig:ddstate}.
	The amplitude of the state $\ket{1010}$, for example, is obtained by multiplying the root weight with the weights along the path alternating between the right and the left successor,~i.e., by multiplying $1 * \tfrac{1}{\sqrt{2}} * \tfrac{1}{\sqrt{2}} * -\tfrac{1}{\sqrt{2}} * \tfrac{1}{\sqrt{2}} = -\tfrac{1}{4} $.
\end{example}

A similar decomposition can be employed for the matrices representing quantum gates by 
recursively splitting the respective matrix into four equally-sized sub-matrices according to the operator basis 
\[
\{\begin{bNiceMatrix}[small]1&0\\0&0\end{bNiceMatrix}, \begin{bNiceMatrix}[small]0&1\\0&0\end{bNiceMatrix}, \begin{bNiceMatrix}[small]0&0\\1&0\end{bNiceMatrix}, \begin{bNiceMatrix}[small]0&0\\0&1\end{bNiceMatrix}\} =
\{\ket{0}\bra{0}, \ket{0}\bra{1}, \ket{1}\bra{0}, \ket{1}\bra{1}\}.
\]

\begin{example}\label{ex:opdd}
	Consider again the operations shown on the left-hand side of \autoref{fig:circuit}. The corresponding decision diagrams are shown in \autoref{fig:ddop}.
\end{example}

As described above, applying a gate to a quantum system entails the matrix-vector multiplication of the corresponding matrix with the current state vector.
This operation can be recursively broken down according to
\begin{align*}
\begin{bmatrix}
U_{00} & U_{01} \\
U_{10} & U_{11} \\
\end{bmatrix}
\cdot
\begin{bmatrix}
\alpha_{0x} \\
\alpha_{1x} \\
\end{bmatrix}
=
\begin{bmatrix}
(U_{00} \cdot \alpha_{0x} + U_{01} \cdot \alpha_{1x}) \\
(U_{10} \cdot \alpha_{0x} + U_{11} \cdot \alpha_{1x}) \\
\end{bmatrix},
\end{align*}
with $U_{ij}\in\mathbb{C}^{2^{n-1}\times 2^{n-1}}$ and $\alpha_{ix}\in\mathbb{C}^{2^{n-1}}$ for $i,j\in\{0,1\}$.
Since the $U_{ij}$ and $\alpha_{ix}$ directly correspond to the successors in the respective decision diagrams, matrix-vector multiplication is a native operation on decision diagrams and its complexity scales with the product of the number of nodes of both decision diagrams.
Thus, whenever the decision diagrams remain compact throughout the computation, the simulation of quantum circuits can be efficiently conducted using decision diagrams~\mbox{\cite{zulehnerAdvancedSimulationQuantum2019, samoladasImprovedBDDAlgorithms2008, viamontesHighperformanceQuIDDBasedSimulation2004, hillmichAccurateNeededEfficient2020, zulehnerMatrixVectorVsMatrixMatrix2019,grurlArraysVsDecision2020}}.

\section{Motivation and General Idea}
\label{sec:general_idea}
Decision diagrams offer a complementary approach for the simulation of quantum circuits.
In many cases, they have been shown to compactly represent and efficiently manipulate the state of a quantum system. However, there remain some obstacles and limitations of decision diagram-based simulation. In the following, we will discuss these limitations and provide the general idea for overcoming them.

\subsection{Limitations of Decision Diagram-based Simulation}
\label{sec:simulation}
While decision diagrams frequently allow to compactly represent the state of a quantum system, in the worst case, their size is still exponential with respect to the number of qubits.
Such situations arise when no redundancy in the description of the quantum state can be exploited and, thus, only a few nodes can be shared.
This occurs, e.g., during the simulation of quantum circuits whose gates are chosen randomly according to some scheme (e.g., the circuits used by Google in their quantum supremacy experiment~\cite{boixoCharacterizingQuantumSupremacy2018, boixoQuestionQuantumSupremacy2018}).
In general, such circuits are designed to make classical simulations as hard as possible, which---in case of decision diagrams---implies that they try hard to not give rise to a particular structure in the corresponding states.

At the same time, decision diagram operations such as matrix-vector multiplication, addition, inner product computation, or sampling, scale polynomially with the number of nodes of the involved decision diagrams.
As such, they are highly efficient whenever the underlying decision diagrams actually emit a compact structure.
However, if the number of nodes in the decision diagram grows exponentially (which happens in the worst case), their performance degrades significantly.
Even worse: In this regime, decision diagrams actually perform worse than, e.g., array-based techniques, which always incur this exponential (memory) complexity, but have a lower overhead of maintaining the underlying data structure.

In one way or another, all Schrödinger-style methods (such as arrays, tensor trains, decision diagrams, etc.) face the problem of exponentially increasing simulation complexity.
Many established simulation methods compensate for this limitation by heavily employing parallelization, i.e., making use of the many cores in today's systems or even large clusters of supercomputers to speed up the computation~\mbox{\cite{hanerPetabyteSimulation45Qubit2017,doiQuantumComputingSimulator2019,jonesQuESTHighPerformance2018,guerreschiIntelQuantumSimulator2020, wuFullstateQuantumCircuit2019}}.
While similar efforts have been conducted towards parallelizing decision diagram-based simulation, e.g., in~\cite{hillmichConcurrencyDDbasedQuantum2020}, no \mbox{\enquote{break-through}} has been achieved there yet.
The main obstacles in this regard are that the shared nature of decision diagrams necessitates inter-process-communication and some kind of locked access to its central data members (such as unique or compute tables) which, in turn, quickly eliminate the benefits of parallelization.

In this work, we entertain an approach that drastically reduces the exponential simulation complexity for certain classes of problems (specifically, depth-limited circuits) at the expense of simulating multiple, independent instances whose contributions are eventually accumulated.
Coincidentally, this further allows to fully utilize all available processing power---effectively \enquote{killing both birds with one stone}.
The proposed approach follows the concepts of a \emph{hybrid \mbox{Schrödinger-Feynman} technique}, which is reviewed next before the general idea of applying this concept to decision diagrams is described.

\pagebreak

\subsection{Hybrid Schrödinger-Feynman Simulation}
\label{sec:slicing}

The hybrid Schrödinger-Feynman simulation style aims at reducing the complexity 
of the Schrödinger-style simulation 
by breaking it down into simpler parts.
This is accomplished by employing concepts from Feynman-style path summation. 
To this end, the most important concept is the Schmidt decomposition of a \mbox{two-qubit} gate:
Any two-qubit gate (represented by a unitary $4\times 4$ matrix~$U$) can be decomposed into at most four tensor products according to
\[ 
U = \ket{0}\bra{0}\otimes U_{00} +  \ket{0}\bra{1}\otimes U_{01} + \ket{1}\bra{0}\otimes U_{10} + \ket{1}\bra{1}\otimes U_{11},
\]
with unitary $U_{ij}\in\mathbb{C}^{2\times 2}$ and $i,j=0,1$.

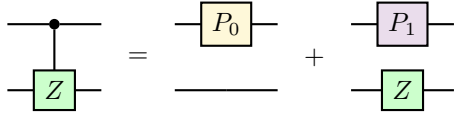
\begin{figure}[t]
	\centering
	\begin{tikzpicture}
	\node[] (G) {
		\begin{quantikz}[column sep=10pt, row sep={12.5pt,between origins}, scale=0.95]
		& \ctrl{2}  & \qw & & & \gate[style={fill=Goldenrod!20}]{P_0} & \qw & & & \gate[style={fill=Orchid!20}]{P_1} & \qw \\
		& & & = & & & & + & & & \\
		& \gate[style={fill=green!20}]{Z} & \qw & & & \qw & \qw & & & \gate[style={fill=green!20}]{Z} & \qw
		\end{quantikz}
	};
	\end{tikzpicture}
	\caption{Schmidt Decomposition of the controlled-Z Gate}
	\label{fig:schmidt}
\end{figure}

\begin{example}\label{ex:schmidt}
	Consider the controlled-Z gate whose unitary matrix representation is given by \mbox{$U=\mbox{diag}(1,1,1,-1)$}. 
	Intuitively, the operation leaves both qubits untouched whenever the control qubit is in state $\ket{0}$, while it applies a $Z$ gate to the target qubit in case the control qubit is in state $\ket{1}$. In formulas:
	\[
	\ket{0}\otimes\ket{x}\mapsto\ket{0}\otimes\mathbb{I}\ket{x} \mbox{ and } \ket{1}\otimes\ket{x}\mapsto\ket{1}\otimes Z\ket{x} \mbox{ for } x=0,1.
	\]
	Consequently, its Schmidt decomposition consists of two terms and is given by
	\[
	U = \ket{0}\bra{0}\otimes \mathbb{I} + \ket{1}\bra{1}\otimes Z = P_0\otimes\mathbb{I} + P_1\otimes Z,
	\]
	with $P_0$ ($P_1$) denoting the projection onto \ket{0} (\ket{1}). 
	This is illustrated in \autoref{fig:schmidt}.
\end{example}

The Schmidt decomposition allows to split the application of any two-qubit gate into separate parts that can be calculated independently. 
As such, each decomposed gate increases the number of simulations (i.e., the runtime) by the amount of factors in its decomposition.
After all individual contributions have been computed, they have to be summed up in order to obtain the full result.

Hybrid Schrödinger-Feynman approaches (such as~\mbox{\cite{aaronsonComplexityTheoreticFoundationsQuantum2016, chen64QubitQuantumCircuit2018, markovQuantumSupremacyBoth2018, pednaultParetoefficientQuantumCircuit2020}}) horizontally partition the whole circuit into blocks by introducing \emph{cuts} through the circuit's list of qubits.
Only cross-block gates, i.e. gates acting across such a cut, are decomposed according to their Schmidt decomposition.
This way, individual blocks are independent from each other and, thus, can be simulated separately. 
Then, the total number of necessary simulations depends on the number of cross-block gates.
As this dependence is exponential in the number of gates (e.g., doubling on each cross-block controlled-Z gate), 
such techniques are most efficient for depth-limited circuits.
However, this still constitutes a large class of quantum algorithms---especially those targeted at near-term quantum computers, which are inherently depth-limited due to noise.

\subsection{General Idea}
\label{sec:slicing_gen_idea}
While decision diagrams offer a complementary approach to quantum circuit simulation that (exponentially) outperforms, e.g., array-based techniques, whenever the number of nodes only grows polynomially, their performance is significantly worse for \enquote{hard} instances (where almost no redundancy can be exploited).
The Schmidt decomposition, as introduced above, allows to reduce the complexity of individual simulations by splitting the circuit into independent blocks that are significantly easier to simulate at the expense of runtime. This concept can readily be applied to decision diagrams.

\begin{example}\label{ex:schmidtdd}
	Recall the controlled-Z gate and its decision diagram (shown in \autoref{fig:dds}).
	As reviewed in \autoref{sec:ddsim}, the successors of a (matrix) decision diagram node encode its action according to the operator basis $\{P_0, \ket{0}\bra{1}, \ket{1}\bra{0}, P_1\}$.
	In case of the controlled-Z gate, the left-most successor (corresponding to~$P_0$) leads to a node representing the identity operation while the right-most successor (corresponding to~$P_1$) leads to a node representing the $Z$ operation.
	Splitting these contributions into individual decision diagrams yields the decomposition shown in \autoref{fig:schmidtdd}, which precisely resembles 	the Schmidt decomposition of the controlled-Z gate from \autoref{ex:schmidt} (illustrated in \autoref{fig:schmidt}). 
\end{example}

Overall, these techniques promise to overcome both obstacles raised above: 
By drastically reducing the complexity of individual simulations, the efficiency of decision diagrams can be fully exploited.
Additionally, no inter-process-communication or locked access is necessary when performing the simulations in parallel since they are independent of another.

Yet, hybrid Schrödinger-Feynman approaches have not been applied to decision diagrams.
Some even believe that realizing such \enquote{circuit cutting techniques} with decision diagrams (as they are proposed in this work) is not possible at all~\cite{hongTensorNetworkBased2020}\footnote{In contrast to quantum circuit simulation as considered in this work,~\cite{hongTensorNetworkBased2020} seeks for a complete representation of a quantum circuit's functionality. Both tasks are related as the functionality of a quantum circuit is obtained from consecutive matrix-matrix multiplication of the individual gate descriptions. Consequently, the results in this work are also applicable in the scenario discussed in~\cite{hongTensorNetworkBased2020}.}.
In the remainder of this paper we show that (1)~this is indeed possible, (2)~which problems arise in the realization, and (3)~how they can be handled.

\section{Decision Diagram-based \\Schrödinger-Feynman Simulation}
\label{sec:slicing_dds}

Following the general idea outlined above potentially allows to overcome the limitations of decision diagram-based quantum circuit simulation discussed in \autoref{sec:simulation}.
In this section, we describe the realization of a hybrid \mbox{Schrödinger-Feynman} technique for decision diagrams and discuss the main bottleneck of the resulting scheme.
Afterwards, we show how this bottleneck can be addressed by relying on decision diagrams where they are most efficient, while leaving the rest to more suitable techniques.

\subsection{Realization}\label{sec:realization}

\begin{figure}[t]
	\centering
	\resizebox{0.8\linewidth}{!}{
	\begin{tikzpicture}
	\node[label={[xshift=0.5cm, yshift=-0.5cm]$CZ$}] (G) {
		\usebox{\czgate}
	};

	\node[right = 0.01cm of G] (EQ) {
		$=$
	};
	
	\node[right = 0.2cm of EQ] (P0xI) {
		$\otimes$
	};	
	\node[above = 0.05cm of P0xI,label={[xshift=0.3cm, yshift=-0.5cm]$P_0$}, rectangle, draw, color=Goldenrod] (P0) {
		\usebox{\pzero}
	};
	\node[below = 0.05cm of P0xI,label={[xshift=0.35cm, yshift=-0.5cm]$I$}, rectangle, draw] (I) {
		\usebox{\identity}
	};	

	\node[right = 0.65cm of P0xI] (PLUS) {
		$+$
	};

	\node[right = 0.65cm of PLUS] (P1xZ) {
		$\otimes$
	};	
	\node[above = 0.05cm of P1xZ,label={[xshift=0.3cm, yshift=-0.5cm]$P_1$}, rectangle, draw, color=Orchid] (P1) {
		\usebox{\pone}
	};
	\node[below = 0.05cm of P1xZ, xshift=0.05cm,label={[xshift=0.35cm, yshift=-0.5cm]$Z$}, rectangle, draw, color=Green] (Z) {
		\usebox{\z}
	};	
		\end{tikzpicture}}
	\caption{Schmidt decomposition of the controlled-Z gate decision diagram}
	\label{fig:schmidtdd}
\end{figure}
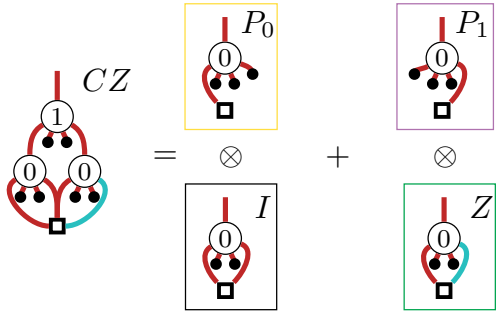

\begin{figure*}
	\centering
	\resizebox{\linewidth}{!}{
	\begin{tikzpicture}
	\pgfmathsetmacro {\yoffseta}{5cm}
	\pgfmathsetmacro {\yoffsetb}{2.5cm}
	\pgfmathsetmacro {\yoffsetc}{1.3cm}

	\node[scale=0.5] (G) {
		\begin{quantikz}[column sep=10pt, row sep={17.5pt,between origins}, scale=0.9]
		\lstick{$q_3$} & \gate[style={fill=RoyalBlue!20}]{H}  & \ctrl{3} & \qw      & \qw \\
		\lstick{$q_2$} & \gate[style={fill=RoyalBlue!20}]{H}  & \qw & \ctrl{3} & \qw \\
		\wave&&&&\\
		\lstick{$q_1$} & \gate[style={fill=RoyalBlue!20}]{H}  & \gate[style={fill=green!20}]{Z} & \qw      & \qw \\
		\lstick{$q_0$} & \gate[style={fill=RoyalBlue!20}]{H}  & \qw      & \gate[style={fill=green!20}]{Z} & \qw
		\end{quantikz}
	};

	\node[scale=0.5,right = 0.3cm of G, yshift=\yoffseta] (D0A) {
			\begin{quantikz}[column sep=10pt, row sep={17.5pt,between origins}, scale=0.9]
			\lstick{$q_3$} & \gate[style={fill=RoyalBlue!20}]{H}  & \gate[style={fill=Goldenrod!20}]{P_0} & \qw      & \qw \\
			\lstick{$q_2$} & \gate[style={fill=RoyalBlue!20}]{H}  & \qw          & \ctrl{3} & \qw \\
			\wave&&&&\\
			\lstick{$q_1$} & \gate[style={fill=RoyalBlue!20}]{H}  & \qw          & \qw      & \qw \\
			\lstick{$q_0$} & \gate[style={fill=RoyalBlue!20}]{H}  & \qw          & \gate[style={fill=green!20}]{Z} & \qw
			\end{quantikz}
	};
	\node[scale=0.5,right = 0.3cm of G, yshift=-\yoffseta] (D0B) {
			\begin{quantikz}[column sep=10pt, row sep={17.5pt,between origins}, scale=0.9]
			\lstick{$q_3$} & \gate[style={fill=RoyalBlue!20}]{H}  & \gate[style={fill=Orchid!20}]{P_1} & \qw      & \qw \\
			\lstick{$q_2$} & \gate[style={fill=RoyalBlue!20}]{H}  & \qw & \ctrl{3} & \qw \\
			\wave&&&&\\
			\lstick{$q_1$} & \gate[style={fill=RoyalBlue!20}]{H}  & \gate[style={fill=green!20}]{Z} & \qw      & \qw \\
			\lstick{$q_0$} & \gate[style={fill=RoyalBlue!20}]{H}  & \qw      & \gate[style={fill=green!20}]{Z} & \qw
			\end{quantikz}
	};

	\node[scale=0.5,right = 0.5cm of D0A, yshift=\yoffsetb] (D1A) {
			\begin{quantikz}[column sep=10pt, row sep={18pt,between origins}, scale=0.9]
			\lstick{$q_3$} & \gate[style={fill=RoyalBlue!20}]{H}  & \gate[style={fill=Goldenrod!20}]{P_0} & \qw \\
			\lstick{$q_2$} & \gate[style={fill=RoyalBlue!20}]{H}  & \gate[style={fill=Goldenrod!20}]{P_0} & \qw \\
			\wave&&&\\
			\lstick{$q_1$} & \gate[style={fill=RoyalBlue!20}]{H}  & \qw       & \qw \\
			\lstick{$q_0$} & \gate[style={fill=RoyalBlue!20}]{H}  & \qw       & \qw
			\end{quantikz}
	};	
	\node[scale=0.5,right = 0.5cm of D0A, yshift=-\yoffsetb] (D1B) {
			\begin{quantikz}[column sep=10pt, row sep={18pt,between origins}, scale=0.9]
			\lstick{$q_3$} & \gate[style={fill=RoyalBlue!20}]{H}  & \gate[style={fill=Goldenrod!20}]{P_0} & \qw \\
			\lstick{$q_2$} & \gate[style={fill=RoyalBlue!20}]{H}  & \gate[style={fill=Orchid!20}]{P_1} & \qw \\
			\wave&&&\\
			\lstick{$q_1$} & \gate[style={fill=RoyalBlue!20}]{H}  & \qw      & \qw \\
			\lstick{$q_0$} & \gate[style={fill=RoyalBlue!20}]{H}  & \gate[style={fill=green!20}]{Z} & \qw
			\end{quantikz}
	};
	\node[scale=0.5,right = 0.5cm of D0B, yshift=\yoffsetb] (D1C) {
			\begin{quantikz}[column sep=10pt, row sep={18pt,between origins}, scale=0.9]
			\lstick{$q_3$} & \gate[style={fill=RoyalBlue!20}]{H}  & \gate[style={fill=Orchid!20}]{P_1} & \qw \\
			\lstick{$q_2$} & \gate[style={fill=RoyalBlue!20}]{H}  & \gate[style={fill=Goldenrod!20}]{P_0} & \qw \\
			\wave&&&\\
			\lstick{$q_1$} & \gate[style={fill=RoyalBlue!20}]{H}  & \gate[style={fill=green!20}]{Z} & \qw \\
			\lstick{$q_0$} & \gate[style={fill=RoyalBlue!20}]{H}  & \qw      & \qw
			\end{quantikz}
	};	
	\node[scale=0.5,right = 0.5cm of D0B, yshift=-\yoffsetb] (D1D) {
			\begin{quantikz}[column sep=10pt, row sep={18pt,between origins}, scale=0.9]
			\lstick{$q_3$} & \gate[style={fill=RoyalBlue!20}]{H}  & \gate[style={fill=Orchid!20}]{P_1} & \qw \\
			\lstick{$q_2$} & \gate[style={fill=RoyalBlue!20}]{H}  & \gate[style={fill=Orchid!20}]{P_1} & \qw \\
			\wave&&&\\
			\lstick{$q_1$} & \gate[style={fill=RoyalBlue!20}]{H}  & \gate[style={fill=green!20}]{Z} & \qw \\
			\lstick{$q_0$} & \gate[style={fill=RoyalBlue!20}]{H}  & \gate[style={fill=green!20}]{Z} & \qw
			\end{quantikz}
	};

	\node[scale=0.5,right = 2cm of D1A, yshift=\yoffsetc] (D2A) {
		\usebox{\dda}
	};
	\node[scale=0.5,right = 2cm of D1A, yshift=-\yoffsetc] (D2B) {
		\usebox{\ddb}
	};
	\node[scale=0.5,right = 2cm of D1B, yshift=\yoffsetc] (D2C) {
		\usebox{\ddc}
	};
	\node[scale=0.5,right = 2cm of D1B, yshift=-\yoffsetc] (D2D) {
		\usebox{\ddd}
	};
	\node[scale=0.5,right = 2cm of D1C, yshift=\yoffsetc] (D2E) {
		\usebox{\dde}
	};
	\node[scale=0.5,right = 2cm of D1C, yshift=-\yoffsetc] (D2F) {
		\usebox{\ddf}
	};
	\node[scale=0.5,right = 2cm of D1D, yshift=\yoffsetc] (D2G) {
		\usebox{\ddg}
	};
	\node[scale=0.5,right = 2cm of D1D, yshift=-\yoffsetc] (D2H) {
		\usebox{\ddh}
	};

	\node[right = 3cm of D1A] (D3A) {
		$\otimes$
	};
	\node[right = 3cm of D1B] (D3B) {
		$\otimes$
	};
	\node[right = 3cm of D1C] (D3C) {
		$\otimes$
	};
	\node[right = 3cm of D1D] (D3D) {
		$\otimes$
	};

	\node[scale=0.5,right = 0.5cm of D3A] (D4A) {
		\usebox{\ddtensora}
	};
	\node[scale=0.5,right = 0.5cm of D3B] (D4B) {
		\usebox{\ddtensorb}
	};
	\node[scale=0.5,right = 0.5cm of D3C] (D4C) {
		\usebox{\ddtensorc}
	};
	\node[scale=0.5,right = 0.5cm of D3D] (D4D) {
		\usebox{\ddtensord}
	};

	\node[right = 8cm of D0A] (D5A) {
		$+$
	};
	\node[right = 8cm of D0B] (D5B) {
		$+$
	};

	\node[scale=0.5,right = 1cm of D5A] (D6A) {
		\usebox{\ddtensoradda}
	};
	\node[scale=0.5,right = 1cm of D5B] (D6B) {
		\usebox{\ddtensoraddb}
	};

	\node[right = 13cm of G] (D7A) {
		$+$
	};
	\node[scale=0.7,right = 0.5cm of D7A] (D8A) {
		\usebox{\ddfullstate}
	};

	\draw [->] (G.45) -- node[above,xshift=-0.1cm] {$0$} (D0A);
	\draw [->] (G.-45) -- node[below,xshift=-0.1cm] {$1$} (D0B);
	
	\draw [->] (D0A.35) -- node[above,xshift=-0.1cm] {$00$} (D1A.180);
	\draw [->] (D0A.-35) -- node[below,xshift=-0.1cm] {$01$} (D1B.180);
	\draw [->] (D0B.35) -- node[above,xshift=-0.1cm] {$10$} (D1C.180);
	\draw [->] (D0B.-35) -- node[below,xshift=-0.1cm] {$11$} (D1D.180);
	
	\draw [->] (D1A.37) -- (D2A);
	\draw [->] (D1A.-37) -- (D2B);
	\draw [->] (D1B.37) -- (D2C);
	\draw [->] (D1B.-37) -- (D2D);
	\draw [->] (D1C.37) -- (D2E);
	\draw [->] (D1C.-37) -- (D2F);
	\draw [->] (D1D.37) -- (D2G);
	\draw [->] (D1D.-37) -- (D2H);

	\draw [->] (D2A) -- (D3A);
	\draw [->] (D2B) -- (D3A);
	\draw [->] (D2C) -- (D3B);
	\draw [->] (D2D) -- (D3B);
	\draw [->] (D2E) -- (D3C);
	\draw [->] (D2F) -- (D3C);
	\draw [->] (D2G) -- (D3D);
	\draw [->] (D2H) -- (D3D);
	
	\draw [->] (D3A) -- (D4A);
	\draw [->] (D3B) -- (D4B);
	\draw [->] (D3C) -- (D4C);
	\draw [->] (D3D) -- (D4D);
	
	\draw [->] (D4A) -- (D5A);
	\draw [->] (D4B) -- (D5A);
	\draw [->] (D4C) -- (D5B);
	\draw [->] (D4D) -- (D5B);
	
	\draw [->] (D5A) -- (D6A);
	\draw [->] (D5B) -- (D6B);
	
	\draw [->] (D6A) -- (D7A);
	\draw [->] (D6B) -- (D7A);

	\draw [->] (D7A) -- (D8A);

	\node[right of = G, yshift=5.5cm] (Label1) {
		$1^{\mathit{st}}$ Decision
	};
	\node[below = 3cm of Label1] (BottomLabel1) {};
	
	\node[right of = D0A, xshift=0.3cm,yshift=3cm] (Label2) {
		$2^{\mathit{nd}}$ Decision
	};
	\node[below = 1cm of Label2] (BottomLabel2) {};
	\node[right of = D1A, xshift=0.6cm, yshift=1.75cm] (Label3) {
		Actual Simulation
	};
	\node[below = 0.5cm of Label3] (BottomLabel3) {};
	\node[left of = D3A, xshift=1cm, yshift=1.75cm] (Label4) {
		Kronecker
	};
	\node[below = 1cm of Label4] (BottomLabel4) {};
	\node[right of = D4A, xshift=0.7cm, yshift=1.75cm] (Label5) {
		$1^{\mathit{st}}$ Additions
	};
	\node[below = 2cm of Label5] (BottomLabel5) {};
	\node[right of = D6A, xshift=-0.15cm, yshift=3cm] (Label6) {
		$2^{\mathit{nd}}$ Addition
	};
	\node[below = 4cm of Label6] (BottomLabel6) {};

	\draw [->] (Label1) -- (BottomLabel1);
	\draw [->] (Label2) -- (BottomLabel2);
	\draw [->] (Label3) -- (BottomLabel3);
	\draw [->] (Label4) -- (BottomLabel4);
	\draw [->] (Label5) -- (BottomLabel5);
	\draw [->] (Label6) -- (BottomLabel6);
	
	\node[below = 0.001 cm of D2H] (Legend) {
		\usebox{\legend}
	};

	\end{tikzpicture}}
	\caption{Hybrid Schrödinger-Feynman simulation of the circuit shown in \autoref{fig:circuit}.}
	\label{fig:slicing}
\end{figure*}
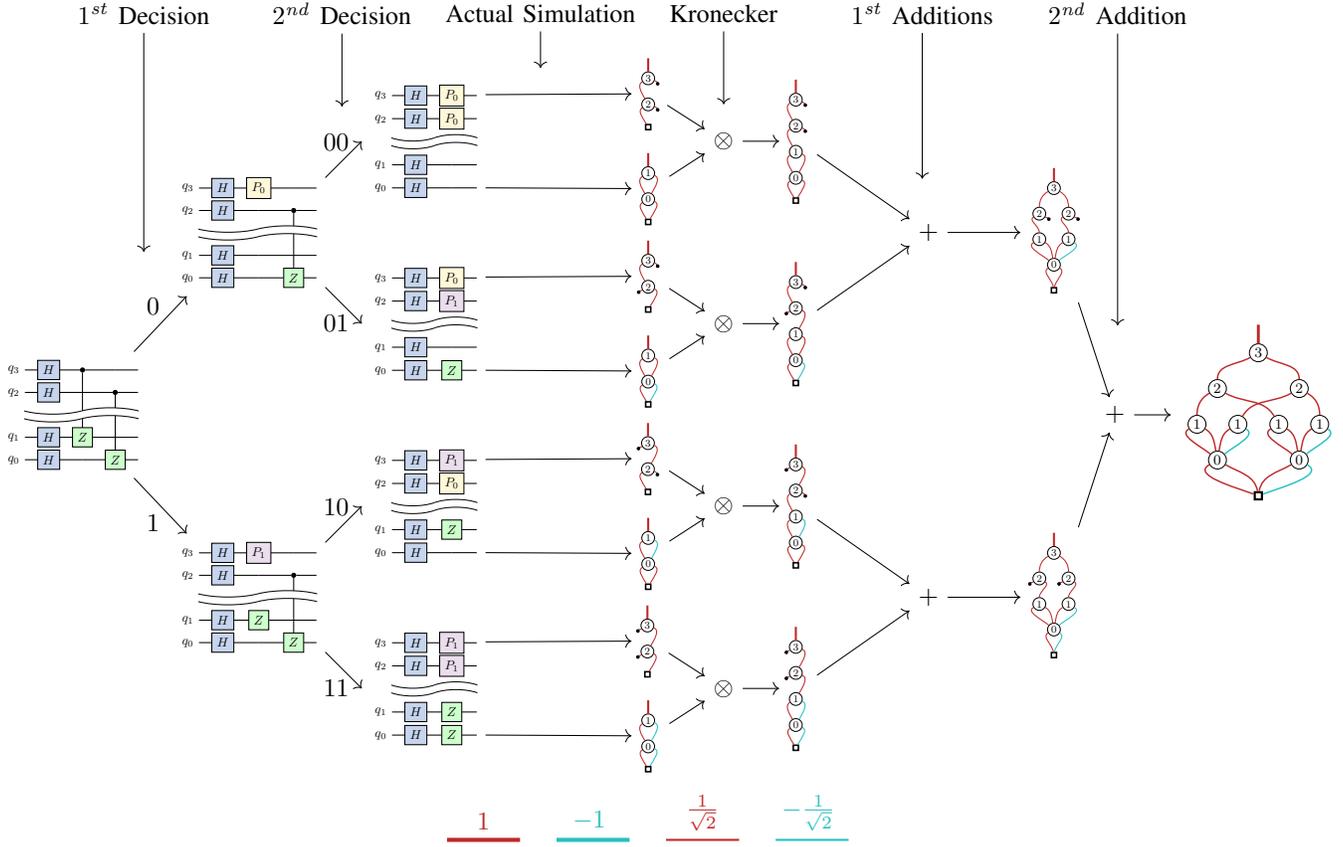

In order to employ a hybrid Schrödinger-Feynman approach, the circuit first has to be \emph{partitioned} into blocks as reviewed in \autoref{sec:slicing}.
In general, there is a large degree of freedom for how to choose such a partitioning, e.g., the number of gates acting across individual blocks (\emph{\mbox{cross-block} gates}).
In the following, we consider splitting the circuit into two (almost) equally-sized blocks---ensuring that each block approximately has the same number of qubits.
Then, the number of simulations to be performed depends on the number of cross-block gates that need to be decomposed (according to their Schmidt decomposition).
Each individual simulation can be assigned a unique identifier that specifies the \emph{decision} (i.e., part of the Schmidt decomposition)  to make at each \emph{decision point} (i.e., cross-block gate).

\begin{example}\label{ex:partitioning}
	Consider again the circuit shown in \autoref{fig:circuit} and assume it is partitioned into two equally-sized halves.
	Then, both controlled-Z gates act across the blocks and, hence, need to be decomposed.
	This yields two decision points with two choices each (the parts of the controlled-Z gate's Schmidt decomposition)---for a total of four individual parts to be simulated as illustrated on the left-hand side of~\autoref{fig:slicing}.
	To this end, the first (second) term of the Schmidt decomposition is encoded as $0$ ($1$).
	Therefore, each simulation can be assigned a bitstring of length two (i.e., number of decisions) that indicates which term of the decomposition is to be calculated.
\end{example}

Splitting the circuit in this fashion has three major benefits. 
Most importantly, the blocks in each individual simulation are independent from another and, thus, can be simulated separately.
This reduces the complexity for each simulation from simulating one $n$-qubit circuit to simulating two \mbox{$n/2$-qubit} circuits---an exponential improvement since the state of an $n$-qubit system grows as $2^n$.
Furthermore, as shown in \autoref{fig:schmidtdd}, the decision diagrams for individual parts of a gate's Schmidt decomposition are typically much less complex than the full decision diagram---allowing for more compact decision diagrams throughout the simulations.
Finally, all individual simulations are independent from another and, hence, can be more efficiently simulated in parallel---even with decision diagrams.

\begin{example}\label{ex:hybridsim}
	Consider again the scenario from \autoref{ex:partitioning}. Then, each of the four individual simulations requires the simulation of two two-qubit circuits.
	The resulting decision diagrams are shown in the middle of \autoref{fig:slicing}.
	A maximum of two nodes (the best case for two-qubit decision diagrams) is required during each individual simulation.
	In contrast, the Schrödinger-style simulation (see \autoref{ex:statedd}) required the handling of up to nine nodes.
\end{example}

Although this only represents a small example, it already shows the potential that the hybrid Schrödinger-Feynman technique brings to decision diagram-based simulation.
However, the computation is not yet finished.
As described in~\autoref{sec:slicing}, the final result of the simulation is obtained as the sum of all individual parts.
First, the simulation results of each block have to be combined by forming the tensor product of the corresponding states.
Forming the tensor product of two decision diagrams is highly efficient, as it merely requires replacing the terminal node of one decision diagram with the root node of the other.

\begin{example}\label{ex:tensor}
	Forming pairwise tensor products of the decision diagrams resulting from the simulations of the circuit from \autoref{ex:hybridsim} yields four decision diagrams of size four each. These decision diagrams (shown in the middle of \autoref{fig:slicing}) represent the state vectors of the individual simulations.
\end{example}

Finally, all the decision diagrams need to be added up to obtain the decision diagram representing the final state vector. 
As for the simulations, these additions can be computed in parallel without inter-process-communication using a \mbox{tree-like} scheme whose depth corresponds to the logarithm of the number of individual simulations.

\begin{example}\label{ex:add}
	The right-hand side of \autoref{fig:slicing} illustrates the process of adding up the individual contributions in order to obtain the final state vector.
	At the first addition level, two decision diagrams of size six result, while after the final addition, the nine-node decision diagram already seen in the right part of \autoref{fig:dds} results.	
\end{example}

\subsection{Decision Diagram Addition as a Bottleneck} \label{sec:adding_overhead}
The combination of all the individual results, i.e., the addition of all resulting decision diagrams, inevitably builds up a potential bottleneck.
While the complexity of the decision diagrams throughout the individual simulations might be drastically lower than the complexity of the full \mbox{Schrödinger-style} simulation, the final decision diagram obviously remains the same.
Consequently, whenever the final decision diagram grows exponentially, this complexity builds up somewhere along the way of adding up the individual contributions.
Since addition of decision diagrams scales linearly with respect to the number of nodes of both decision diagrams, subsequent additions in the \enquote{addition hierarchy} take longer and longer.

At some point, this constitutes a severe bottleneck for the hybrid Schrödinger-Feynman simulation using decision diagrams because the overhead of maintaining the data structure becomes larger than the benefit gained from a potentially more compact representation.
As confirmed by our experimental evaluations (which are summarized later in \autoref{sec:results}), decision diagrams are highly efficient when it comes to the first part of the hybrid Schrödinger-Feynman scheme (i.e., the simulation of individual parts constituting the overall result), while it might get more challenging in the second part of the computation (i.e., adding up all the individual contributions).
In the following, we show how this bottleneck can be addressed whenever the final decision diagram is going to be exponentially large.

\subsection{Avoiding the Final Overhead}\label{sec:ampadd}

Thus far, we used decision diagrams for the first part of the  hybrid Schrödinger-Feynman scheme because of their efficiency in simulation. In the second part (when the final results are determined by addition), this benefit might disappear and lead to exponentially large decision diagrams. 
At this point, decision diagrams do not offer any advantages anymore compared to simpler data-structures such as arrays (in fact, the overhead caused by maintaining a dedicated data-structure will make decision diagrams perform even worse than arrays, which require practically no overhead).  
That is, in these cases, one is better off extracting the state vector represented by the decision diagram into an array and continue working with that representation.
That way, one relies on decision diagrams where they are most efficient, while substituting more direct representations whenever the limit for decision diagrams has been reached.

To this end, the complete vector represented by a decision diagram is extracted with a single recursive traversal of the decision diagram by accumulating amplitude contributions along the edges.
After the extraction, all resulting arrays can be added together to obtain the final state vector.
As a consequence, one benefits from the memory locality of array-based representations as well as vectorized instruction support of modern CPUs---completely circumventing the overhead decision diagram-based addition incurs in this regime.

\begin{example}\label{ex:ddextract}
	Consider again the hybrid Schrödinger-Feynman scheme shown in \autoref{fig:slicing} for simulating the circuit from \autoref{fig:circuit}.
	After all four individual simulations have been conducted, the amplitudes of the corresponding decision diagrams representing the state vectors are extracted.
	For the top-most decision diagram (corresponding to the \enquote{$00$-decision}) this results in the following (recursive) computation:
	\begin{align*}
		\tfrac{1}{2} * [[\cdots]\; 0000\,0000]^\top
		= \tfrac{1}{2} * [[\cdots]\,0000\; 0000\,0000]^\top \\
		= \tfrac{1}{2} * [[\tfrac{1}{\sqrt{2}}*[\cdots]\,\tfrac{1}{\sqrt{2}}*[\cdots]]\,0000\; 0000\,0000]^\top \\
		= \tfrac{1}{2\sqrt{2}} * [[\tfrac{1}{\sqrt{2}}\; \tfrac{1}{\sqrt{2}}]\,[\tfrac{1}{\sqrt{2}}\; \tfrac{1}{\sqrt{2}}]\,0000\; 0000\,0000]^\top \\
		= \tfrac{1}{4} * [1111\,0000\; 0000\,0000]^\top
	\end{align*}
	Overall, the extraction results in the respective amplitude arrays
	\begin{align*}
	\tfrac{1}{4} * [{\phantom{-}1}{\phantom{-}1}{\phantom{-}1}{\phantom{-}1}{\phantom{-}0}{\phantom{-}0}{\phantom{-}0}{\phantom{-}0}{\phantom{-}0}{\phantom{-}0}{\phantom{-}0}{\phantom{-}0}{\phantom{-}0}{\phantom{-}0}{\phantom{-}0}{\phantom{-}0}\phantom{-}]^\top,\\
		\tfrac{1}{4} * [{\phantom{-}0}{\phantom{-}0}{\phantom{-}0}{\phantom{-}0}{\phantom{-}1}{-1}{\phantom{-}1}{-1}{\phantom{-}0}{\phantom{-}0}{\phantom{-}0}{\phantom{-}0}{\phantom{-}0}{\phantom{-}0}{\phantom{-}0}{\phantom{-}0}\phantom{-}]^\top,\\
		\tfrac{1}{4} * [{\phantom{-}0}{\phantom{-}0}{\phantom{-}0}{\phantom{-}0}{\phantom{-}0}{\phantom{-}0}{\phantom{-}0}{\phantom{-}0}{\phantom{-}1}{\phantom{-}1}{-1}{-1}{\phantom{-}0}{\phantom{-}0}{\phantom{-}0}{\phantom{-}0}\phantom{-}]^\top,\\
		\tfrac{1}{4} * [{\phantom{-}0}{\phantom{-}0}{\phantom{-}0}{\phantom{-}0}{\phantom{-}0}{\phantom{-}0}{\phantom{-}0}{\phantom{-}0}{\phantom{-}0}{\phantom{-}0}{\phantom{-}0}{\phantom{-}0}{\phantom{-}1}{-1}{-1}{\phantom{-}1}\phantom{-}]^\top,
	\end{align*}
	which are subsequently added up to produce the final state vector
	\[
	\tfrac{1}{4}* [{\phantom{-}1}{\phantom{-}1}{\phantom{-}1}{\phantom{-}1}{\phantom{-}1}{-1}{\phantom{-}1}{-1}{\phantom{-}1}{\phantom{-}1}{-1}{-1}{\phantom{-}1}{-1}{-1}{\phantom{-}1}\phantom{-}]^\top.
	\]
\end{example}

As our experimental evaluations (which are summarized next) confirm, employing decision diagrams for the individual simulations of the hybrid Schrödinger-Feynman scheme, while handing off the accumulation of individual results to computations on arrays, allows to capitalize on the best of both worlds and mitigates the limitations discussed in \autoref{sec:realization}.

\begin{table*}[t]
	\sisetup{table-format=>4.2, round-mode=places, round-precision=2, group-minimum-digits=4, table-number-alignment=right}
	\caption{Experimental Evaluations}
	\label{tbl:evaluation_results}	
	\centering
	\resizebox{0.73\linewidth}{!}{
	\begin{tabular}{lr!{\quad}r!{\qquad}rr!{\qquad}>{\bfseries}rr}
			\multicolumn{2}{c}{Benchmark} & \multicolumn{1}{c}{JKQ DDSIM~\cite{zulehnerAdvancedSimulationQuantum2019}} & \multicolumn{2}{c}{DD-based (\autoref{sec:realization})} & \multicolumn{2}{c}{Amp. Adding (\autoref{sec:ampadd})} \\
		\cmidrule(l{2em}r{2em}){1-2}\cmidrule(l{2em}r{2em}){3-3}\cmidrule(l{2em}r{2em}){4-5}\cmidrule(l{2em}r{2em}){6-7}
		Name & {\#Decisions} & {$t_\mathit{ref}$~[\si{\second}]} & {$t_\mathit{DD}$~[\si{\second}]} & {$t_\mathit{ref}/t_\mathit{fullDD}$} & {$t_\mathit{amp}$~[\si{\second}]} & {$t_\mathit{ref}/t_\mathit{amp}$}\\
		\midrule
		inst\_4x4\_10\_0 & 4 & \tablenum{0.349561} & \tablenum{0.277305} & \tablenum{1.260565} & \tablenum{0.038140} & \tablenum{9.1652} \\
		inst\_4x4\_10\_9 & 4 & \tablenum{0.469749} & \tablenum{0.242852} & \tablenum{1.9343} & \tablenum{0.034840} & \tablenum{13.483}\smallskip\\
		inst\_4x4\_14\_0 & 6 & \tablenum{2.293602} & \tablenum{1.371637} & \tablenum{1.672164} & \tablenum{0.131064} & \tablenum{17.4998} \\
		inst\_4x4\_14\_9 & 6 & \tablenum{3.231343} & \tablenum{1.325871} & \tablenum{2.437147} & \tablenum{0.129097} & \tablenum{25.03035}\smallskip\\
		inst\_4x4\_15\_0 & 8 & \tablenum{2.995799} & \tablenum{1.251144} & \tablenum{2.3944478} & \tablenum{0.342764} & \tablenum{8.74012} \\
		inst\_4x4\_15\_9 & 8 & \tablenum{4.187732} & \tablenum{1.375071} & \tablenum{3.045466} & \tablenum{0.368718} & \tablenum{11.35755}\smallskip\\
		inst\_4x4\_19\_0 & 10 & \tablenum{5.153476} & \tablenum{3.007396} & \tablenum{1.71359} & \tablenum{1.360783} & \tablenum{3.78712} \\
		inst\_4x4\_19\_9 & 10 & \tablenum{5.637097} & \tablenum{3.120667} & \tablenum{1.806376} & \tablenum{1.516366} & \tablenum{3.7175}\bigskip\\
		inst\_4x5\_10\_0 & 5 & \tablenum{464.814178} & \tablenum{67.208443} & \tablenum{6.91601} & \tablenum{0.190910} & \tablenum{2434.7293} \\
		inst\_4x5\_10\_9 & 5 & \tablenum{50.246006} & \tablenum{16.244240} & \tablenum{3.093158} & \tablenum{0.198209} & \tablenum{253.500}\smallskip\\
		inst\_4x5\_13\_0 & 8 & \tablenum{1340.926636} & \tablenum{266.139343} & \tablenum{5.03844} & \tablenum{0.682183} & \tablenum{1965.6407} \\
		inst\_4x5\_13\_9 & 8 & \tablenum{1199.967407} & \tablenum{347.682556} & \tablenum{3.45133} & \tablenum{0.808206} & \tablenum{1484.729644}\smallskip\\
		inst\_4x5\_15\_0 & 10 & \tablenum{1859.158203} & \tablenum{277.017090} & \tablenum{6.71134} & \tablenum{2.444238} & \tablenum{760.628958}\\
		inst\_4x5\_15\_9 & 10 & \tablenum{1713.829590} & \tablenum{285.346130} & \tablenum{6.00614} & \tablenum{2.894771} & \tablenum{592.0432}\smallskip\\ 
		inst\_4x5\_20\_0 & 13 & \tablenum{2946.597168} & \tablenum{272.614380} & \tablenum{10.80866} & \tablenum{25.738625} & \tablenum{114.4815}\\
		inst\_4x5\_20\_9 & 13 & \tablenum{2763.026855} & \tablenum{354.490784} & \tablenum{7.7943} & \tablenum{29.463116} & \tablenum{93.77917}\bigskip\\
		inst\_5x5\_10\_0 & 6 & \SI{>24}{\hour} & \SI{>24}{\hour} & --- & \tablenum{3.789385} & --- \\
		inst\_5x5\_10\_9 & 6 & \SI{>24}{\hour} & \SI{>24}{\hour} & --- & \tablenum{4.327647} & ---\smallskip\\
		inst\_5x5\_13\_0 & 10 & \SI{>24}{\hour} & \SI{>24}{\hour} & --- & \tablenum{16.500553} & --- \\
		inst\_5x5\_13\_9 & 10 & \SI{>24}{\hour} & \SI{>24}{\hour} & --- & \tablenum{22.694086} & ---\smallskip\\
		inst\_5x5\_20\_0 & 15 & \SI{>24}{\hour} & \SI{>24}{\hour} & --- & \tablenum{999.358887} & --- \\
		inst\_5x5\_20\_9 & 15 & \SI{>24}{\hour} & \SI{>24}{\hour} & --- & \tablenum{1211.252197} & ---\bigskip\\
		inst\_5x6\_10\_0 & 7 & \SI{>24}{\hour} & \SI{>24}{\hour} & --- & \tablenum{137.146835} & --- \\
		inst\_5x6\_10\_9 & 7 & \SI{>24}{\hour} & \SI{>24}{\hour} & --- & \tablenum{194.918854} & ---\\\bottomrule
	\end{tabular}}
\end{table*}

\section{Experimental Results}
\label{sec:results}

In order to evaluate the effectiveness of decision \mbox{diagram-based} hybrid \mbox{Schrödinger-Feynman} simulation, the scheme proposed in this work has been implemented on top of the state-of-the-art decision diagram-based simulator \mbox{JKQ DDSIM}, which based on~\cite{zulehnerAdvancedSimulationQuantum2019} and is part of the JKQ toolset for quantum computing~\cite{willeJKQJKUTools2020}.
The corresponding implementation is available at \url{https://github.com/iic-jku/ddsim}.

As benchmarks, we considered Google's supremacy circuits~\cite{boixoQuestionQuantumSupremacy2018} which constitute one of the hardest instances for quantum circuit simulation to date and, because they belong to the class of depth-limited circuits, are exactly the kind of circuits for which the proposed scheme is particularly suited for (see \autoref{sec:slicing}).
The evaluations were conducted on machine equipped with a \mbox{$16$-core AMD Ryzen 9 3950X} CPU and \SI{128}{\giga\byte} RAM running Ubuntu 20.04.
\emph{Double} precision floating points and a hard timeout of \SI{24}{\hour} were used for all simulations.

\autoref{tbl:evaluation_results} shows the results of our evaluations.
Here, the first two columns identify the benchmark circuit and the respective number of decisions (i.e., cross-block gates)\footnote{The circuits chosen for this evaluation only use controlled-Z gates as \mbox{two-qubit} gates. Consequently, for a certain number of decisions $x$, $2^x$ individual simulations have to be performed in the hybrid \mbox{Schrödinger-Feynman} scheme.}.
Then, the runtime of the JKQ DDSIM Schrödinger-style simulator is listed, while the remaining four columns contain the runtime and the speedup for the general decision \mbox{diagram-based} hybrid \mbox{Schrödinger-Feynman} scheme (see~\autoref{sec:realization}) and the optimized scheme using arrays for the final additions (see~\autoref{sec:ampadd}), respectively.

From the results, it can be seen that, the higher the number of qubits in the circuit, the higher the potential gain of employing the hybrid Schrödinger-Feynman scheme.
While using the general scheme from \autoref{sec:realization} only yields a small speedup for the smallest benchmarks (i.e. \textit{inst\_4x4\_X\_Y}), the medium-sized benchmarks (i.e. \textit{inst\_4x5\_X\_Y}) show an average speedup of $\approx 6.2\times$. However, neither the JKQ DDSIM \mbox{Schrödinger-style} simulator, nor the general scheme proposed in \autoref{sec:realization}, were able to simulate the larger benchmarks (i.e. \textit{inst\_5x5\_X\_Y} and \textit{inst\_5x6\_X\_Y}) within \SI{24}{\hour}.
As discussed in \autoref{sec:adding_overhead}, this can be attributed to the fact that decision diagram addition on exponentially growing decision diagrams poses a severe bottleneck.

This problem is addressed by using decision diagrams for the individual simulations and resorting to arrays for the final additions. In fact, the numbers confirm that, then, speedups of several factors and up to several orders of magnitude can be observed across all benchmarks. More impressively, even the biggest circuits in our evaluations that could not be simulated in a whole day using the \mbox{Schrödinger-style} simulator can be simulated in roughly \SI{20}{\minute} using this scheme.

\section{Conclusions}
\label{sec:conclusions}
In this work, we showed that a hybrid Schrödinger-Feynman technique can be applied to decision diagram-based quantum circuit simulation, which, for the first time, allows to fully exploit the available hardware resources.
Due to the substantially decreased complexity of the individual simulations, decision diagrams are employed in a regime where more redundancy can be exploited.
By handing off the accumulation of individual results to computations on arrays, the bottleneck of decision diagram addition observed in this work can be effectively circumvented.
The resulting scheme combines the best of both worlds and allows to significantly advance the state of the art in decision diagram-based quantum circuit simulation.
An implementation of the proposed simulation technique is publicly available at \url{https://github.com/iic-jku/ddsim}.

\section*{Acknowledgments}
This project has received funding from the European Research Council (ERC) under the European Union’s Horizon 2020 research and innovation programme (grant agreement No. 101001318).
It has partially been supported by the LIT Secure and Correct Systems Lab funded by the State of Upper Austria as well as by the BMK, BMDW, and the State of Upper Austria in the frame of the COMET program (managed by the FFG).

\printbibliography
\end{document}